\documentclass{jaa}
\usepackage{graphicx}
\usepackage{caption}
\usepackage{subcaption}
\usepackage{hyperref}
\begin{document}\sloppy

\title{AstroSat-CZTI as a hard X-ray Pulsar Monitor}

\author{Anusree, K.G.\textsuperscript{1,*}, Bhattacharya, D.\textsuperscript{2}, Rao, A.R.\textsuperscript{2,5}, Vadawale, S.\textsuperscript{3}, Bhalerao,  V.\textsuperscript{4} \and Vibhute, A.\textsuperscript{2}. }
\affilOne{\textsuperscript{1}School of Pure and Applied Physics, Mahatma Gandhi University, Kottayam 686560, India.\\}
\affilTwo{\textsuperscript{2}Inter-University Centre for Astronomy and Astrophysics, Pune 411007, India.\\}
\affilThree{\textsuperscript{3}Physical Research Laboratory, Ahmedabad, Gujarat 380009, India.\\}
\affilFour{\textsuperscript{4}Indian Institute of Technology, Bombay 400076, India.\\}
\affilFive{\textsuperscript{5}Tata Institute of Fundamental Research, Mumbai 400005, India}
\twocolumn[{
\maketitle
\corres{anusreeaz@gmail.com}

\msinfo{30 October 2020}{18 January 2021}

\begin{abstract}
    The Cadmium Zinc Telluride Imager (CZTI) is an imaging instrument onboard AstroSat. This instrument operates as a nearly open all-sky detector above ~60 keV, making possible long integrations irrespective of the spacecraft pointing. We present a technique based on the AstroSat-CZTI data to explore the hard X-ray characteristics of the $\gamma$-ray pulsar population. We report highly significant ($\sim 30\sigma$) detection of hard X-ray (60--380 keV) pulse profile of the Crab pulsar using $\sim$5000 ks of CZTI observations within 5 to 70 degrees of Crab position in the sky, using a custom algorithm developed by us. Using Crab as our test source, we estimate the off-axis sensitivity of the instrument and establish AstroSat-CZTI as a prospective tool in investigating hard X-ray characteristics of $\gamma$-ray pulsars as faint as 10 mCrab.

\end{abstract}

}]

\doinum{12.3456/s78910-011-012-3}
\artcitid{\#\#\#\#}
\volnum{000}
\year{0000}
\pgrange{1--}
\setcounter{page}{1}
\lp{1}

\section{Introduction}

    Pulse profiles of rotation powered pulsars are shaped by the geometry of the emission region as well as the radiation processes at work in pulsar magnetosphere (Watters and Romani, 2011, Pierbattista et al., 2014). The emission is usually broadband, covering radio through $\gamma$-rays. However, emission at different wavebands typically arises in different regions of the magnetosphere, giving rise to pulse profiles that are wavelength dependent. There exist a variety of magnetospheric models (e.g.\ Cheng, Ho \& Ruderman 1986, Muslimov \& Harding 2004, P\'etri 2011, Cerutti \& Beloborodov 2017) which differ in their prediction of the shape and distribution of acceleration and radiation zones.  Broadband Spectral Energy Distribution (SED) and the energy dependence of the pulse shape and arrival time provide clues to the distribution of these zones and have been used to constrain the magnetospheric geometry and to discriminate between theoretical models  (e.g.\ Romani \& Yadigaroglu 1995, Cheng et al., 2000, Abdo et al., 2009c, Abdo et al. 2010c, Bai \& Spitkovsky 2010, YuanJie et al., 2012, Pierbattista et al., 2012, 2014). 
    
    The number of known $\gamma$-ray pulsars stood at just seven until the launch of NASA's Fermi Large Area Telescope (LAT) in 2008. Since then, the continuous accumulation of data, together with highly efficient searching algorithm, has resulted in the number of pulsars detected by LAT in the $\gamma$-ray band to mount to 253 (Ray et al., 2020), of which 71 have no radio counterpart. The modelling of $\gamma$-ray and radio emission together can provide important constraints on the global magnetospheric properties (e.g.\ see P\'etri \& Mitra 2020). For radio-quiet $\gamma$-ray pulsars, the magnetospheric X-ray emission provides the only additional clue to the emission process.  The radiation energy output of a rotation powered pulsar typically peaks in the X-ray/$\gamma$-ray region. However, only 18 pulsars have been detected in the X-ray band till date (Caraveo 2014, Kuiper and Hermsen 2015) and the SED of many of them are sparsely sampled. Using ephemerides determined from LAT data, four of the LAT radio-quiet pulsars have been observed at photon energies $<$20~keV (Lin et al., 2010; Caraveo et al., 2010; Marelli et al., 2014; Lin et al., 2013 and Lin et al., 2014). These fall in the category of ``Geminga-like" pulsars, for which the profile and spectra are known at soft X-rays (i.e., $<$20keV) and $\gamma$-rays alone (PSR B0633+17, Halpern and Holt 1992; Bertsch et al., 1992). Unfortunately, at energies above 20 keV, the major X-ray observatories have had relatively low sensitivity.
    Additional deep observations in the hard X-ray band would therefore augment the existing information, providing a better estimate of spectral shape and bolometric luminosity of these and other pulsars hitherto undetected in X-rays. Three major emission mechanisms operate in pulsar magnetospheres,  namely synchrotron radiation, curvature radiation and inverse-Compton scattering.   A well-measured SED  can distinguish the relative contributions of these components, leading to a model of the particle energy distribution in the emission zones.

    To increase the sample of pulsar detections in the hard X-ray/Soft $\gamma$-ray bands, and to investigate how their properties fit in the general picture emerging from the theoretical studies of the Fermi's young gamma-ray pulsars, we need ``open all-sky X-ray detectors". The pulsar spectra are steep at high energies. In general, photon flux of the young/middle-aged LAT $\gamma$-ray pulsars can be represented by a power-law with a simple exponential cutoff, i.e.\ $F_{\gamma} = k \cdot  (E_{\gamma} /E_{0} )^{\Gamma} \cdot  \exp(-(E_{\gamma} /E_{c} )^{\beta} )$ where $\beta \approx 1$ and $E_{\gamma}$ is the photon energy, $E_{0}$ a normalisation energy, $ E_{c}$ the cutoff energy and $k$ is the normalization. The photon index $\Gamma$ has been found to lie in the range $-0.4$ to $-1.7$ (Kuiper and Hermsen,  2015). Even with a highly sensitive detector with sub-second timing resolution, it takes extremely long exposures to detect a significant number of photons from a typical $\gamma$-ray pulsar. Such long integrations are not affordable by any of the missions at present. Open all-sky detectors, on the other hand, can collect photons during other observations, making it possible to search for these pulsars.

    The first Indian multi-wavelength Satellite AstroSat was launched in 2015 with five instruments onboard (Singh et al., 2014).  One of them, the Cadmium Zinc Telluride Imager (CZTI; Bhalerao et al., 2017), can detect photons in 20--380~keV energy range. Its housing and collimators are made of Aluminium alloy and thin Tantalum shields that define its low-energy field of view but allow sufficient uncollimated penetration above $\sim 60$~keV to make CZTI an excellent wide-angle monitor at higher energies, covering roughly one-third of the sky at all times. This monitoring capability has been leveraged to detect many transients, including over 300 Gamma-Ray Bursts (Sharma et al., this volume). A more detailed description of CZTI can be found in Bhalerao et al., 2017. The aim of this paper is to assess the suitability of CZTI for the detection and study of pulsars in the hard X-ray band, using its off-axis detection capability.

    During CZTI pointing observations, photons from candidate hard X-ray sources that shine in through the walls will come to piggyback on any ongoing observation, with varying sensitivity depending on the pointing.  Arrival times of the photons from a pulsar carry the signature of its spin period, enabling us to search the data for hard X-ray pulsars with known ephemeris. This presents the detection of Crab Pulsar in the energy range 60-380 keV from off-axis CZTI observations, using a custom algorithm developed by us. The Crab pulsar (PSRJ0534+2200) is well-studied at all wavelengths. Using Crab as our calibration source, we determine the off-axis sensitivity of the instrument.

    This paper is structured as follows. Section 2 describes the instruments used in this work and the analysis of data, followed by a discussion of results in section 3 which includes the comparison of hard X-ray pulse profiles of the Crab pulsar obtained from CZTI with $\gamma$-ray profiles from LAT data. Finally, we assess the potential of AstroSat-CZTI in the investigation of hard X-ray counterparts of $\gamma$-ray pulsars.

\section{Data and analysis}
    This work is based on data from AstroSat-CZTI pointing observations that were released for public use on or before 30th April 2019. We have also used publicly available archival data from NASA's Fermi-LAT mission. All material informations about the instruments and data, along with the characteristics of analysis methods, are described in this section. Moreover, we also discuss some of our checks against the vulnerabilities anticipated during long integration.

\begin{table*}[htb]
\tabularfont
\caption{}
\label{table1}
\begin{tabular}{lcccc}
\topline
Telescope& Energy range& Start MJD  & Stop MJD& Exposure(ks)\\\midline
AstroSat-CZTI( Crab at nominal pointing)&30-60 keV&57290&58232&519\\
AstroSat-CZTI(Crab within 5-70 degree of nominal pointing) &60-380 keV&57290&58232&4752\\
Fermi-LAT( within 1 degree of $\gamma$-ray position of Crab  pulsar) &0.1-300 GeV&57290&58500 &5962\\
\hline
\end{tabular}
\tablenotes{Brief summary of the observations. Participating telescopes and their instruments, the energy range chosen for the work, range of MJDs for which the data have been collected for this work and the total exposure achieved are listed}
\end{table*}


\subsection{AstroSat-CZTI} 
    CZTI is the AstroSat instrument primarily designed for simultaneous hard X-ray imaging and spectroscopy of celestial X-ray sources in the energy band 20-150 keV. Its functioning employs the technique of coded mask imaging. The CZTI instrument bears a two-dimensional coded aperture mask (CAM) above its pixellated, 5-mm thick solid-state CZT detector modules spread over four quadrants. Passive collimators are placed in each quadrant of CZTI that support the coded mask. The set up defines a 4.6 deg $\times$ 4.6 deg field of view in $\sim$20-100 keV range with an angular resolution of $\sim$ 8 arcmins. A critical aspect of the collimators and mask is that these are designed to be effective up to $\sim 100$~keV above which they become transparent. The transparency being a function of energy and angle of incidence, appreciable sensitivity for off-axis sources extends down to $\sim 60$~keV (see Bhattacharya et al., 2018).

    The $976cm^{2}$ of the detector's total geometric area is distributed over 16384 pixels, with 4096 pixels in each independent quadrant. After the launch, about 15 percent of the pixels were disabled for having shown excessive electronic noise. Nearly 25 percent of the remaining pixels seemed to have an inadequate spectroscopic response. Considering that the coded mask has a $\sim$ 50\% open fraction, the total effective area at normal incidence is $\sim 420 cm^{2}$ in all active pixels at energies below 100 keV.  The detected events are recorded with a time resolution of 20 $\mu$s. The absolute timestamp assigned to CZTI events is estimated to have a jitter of $\sim 3\,\mu$s RMS (Bhattacharya, 2017) and a fixed offset with respect to Fermi of $650\pm70\,\mu$s (Basu et al., 2018).

\subsubsection{Analysis of CZTI data:}

     All available CZTI pointing observations during MJD 57366-58362, with the Crab pulsar within 5-80 degrees from normal incidence, were selected for this work, resulting in a total exposure of 5096ks. The merged Level-1 data of all the selected CZTI observations were reduced to Level-2 using standard CZTI analysis software. Details and the sequence of analysis modules can be found in the latest version of the CZTI user-guide available at the ASSC website\footnote{http://astrosat-ssc.iucaa.in/}. There are intervals during pointing observation where data is absent due to SAA passage and data transmission loss. There are also intervals when the earth occults the target source. To generate science products from such observations, identifying such intervals and removing the data for that duration by adequately accounting for the gaps is essential. This task is performed by the module {\tt cztgtigen}. It generates Good Time Interval (GTI) files based on various threshold parameters. We generated custom GTI files to take into consideration the earth occultation of the off-axis source and filtered the original event file accordingly. These filtered, clean event files were used in the subsequent steps of analysis.
     
     In this work, we have used AstroSat CZTI observations with on-axis pointing of the Crab pulsar as well as others when the pulsar is off-axis at angles between 5 to 70 deg (see below for the choice of this angle range).  We have also used Fermi-LAT observations of the Crab pulsar spanning a similar time range for comparison.  A brief description of the data sets used is provided in Table \ref{table1}.
     
     The arrival times of the CZTI events at the spacecraft were converted to those at the Solar-system Barycenter using the JPL DE200 Solar-system ephemeris. This was done using the tool {\tt as1bary}, a version of NASA/HEASOFT AXBARY package, customised for AstroSat, using the well-known astrometric position of the Crab pulsar.

     We developed a custom code to fold the barycentered event data spanning many months but including frequent gaps. This code assigns an absolute phase to each recorded event using the polynomial model timing solutions known as SSB polycos generated using tempo2 (Hobbs et al.,2006). Polycos predict a pulsar's parameters at a particular epoch. A polyco file contains pulsar ephemerides over a short period, typically hours, in simple polynomial expansion. All the polycos were generated at an epoch in the centre of each 6-hour interval in this work. These 6-hour ephemerides were then used for folding the LAT data as well as the CZTI data. As an initial test of the custom code, Fermi-LAT $\gamma$-ray events for a few pulsars were folded and compared with their published timing models. The SSB polycos were generated from LAT timing models available publicly at Fermi's website. \footnote{ https://confluence.slac.stanford.edu/display/GLAMCOG/LAT+Gamma-ray+Pulsar+Timing+Models}. Finally, the code was tested on AstroSat-CZTI data. For this,  publicly available AstroSat-CZTI data from six pointing observations of Crab pulsar (Table \ref{table1}, first row) were selected and reduced using the default CZTI data analysis pipeline. The 30-60 keV pulse profile, thus obtained, is shown in Figure \ref{fig2}(a). The Crab pulsar ephemeris derived from Fermi observations and used in this work are given in Table \ref{table2} along with the reference epochs.  

\begin{figure}[t]
\centering
\includegraphics[width=1\columnwidth,height=1\columnwidth,angle=270]{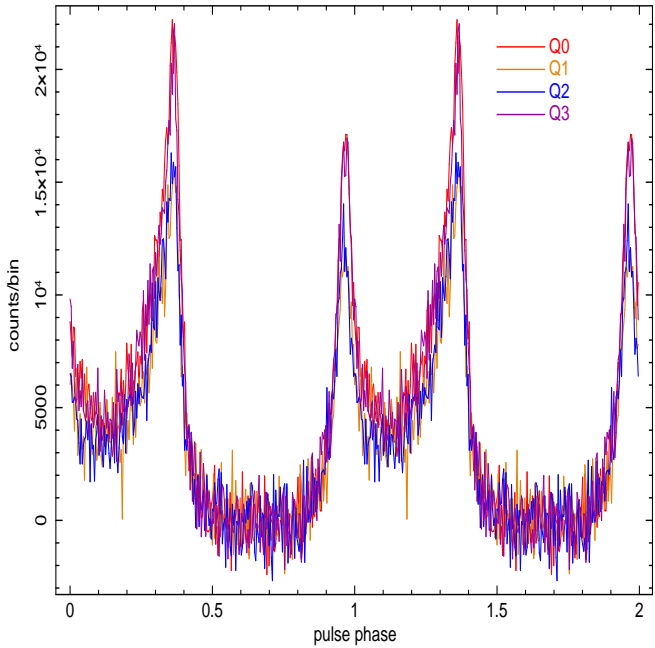}
\caption{Phase histograms of PSR J0534+2200 in 60-380 keV band obtained from 5-70 degree off-angle data (this work) from the four different quadrants (Q0, Q1, Q2 and Q3) of CZTI operating as independent detectors. All the quadrants are found to be time aligned with no measurable relative delay}
\label{fig1}
\end{figure}

    To compare the pulse profiles between hard X-rays and $\gamma$-rays, we folded all the events extracted from AstroSat-CZTI observations and those obtained from Fermi observation within MJD 57290–58500 using the $\gamma$ ray ephemeris mentioned above. For all CZTI data analysis, we have directly combined the data of all four quadrants, as they run on a synchronised time reference (see Figure \ref{fig1}).  We initially folded the off-axis data separately for every 10-degree angle interval and found that when the source is located beyond 70 degrees from the pointing axis, the signal-to-noise ratio in a given integration time drops significantly below those for smaller off-axis angles. We, therefore, decided to restrict the accumulation of off-axis data to the angle range of 5 to 70 deg (below 5 deg the source would appear in the main FoV).
    
    With the phase reference as presented in Table \ref{table3}, we consider the ``off-pulse'' region of the profile, phase 0.5 through 0.8, as the background. The signal to noise ratio is calculated as the ratio of the peak count to the standard deviation of the counts in the off-pulse region. We also divided the obtained X-ray events into several different energy bands to check consistency, as shown in Figure \ref{fig2}(b).

\begin{table}[htb]
\tabularfont
\caption{Fermi-LAT $\gamma$-ray timing solution of PSR J0534+2200 by Fermi Timing Observers. The numbers in parentheses
denote $1\sigma$ errors in the parameters}
\label{table2}. 
\begin{tabular}{cc}
\hline
Parameters&   ~\\\midline
Right ascension,~~ $\alpha$ &05:34:31.94\\
Declination,~~$\delta$ &+22:00:52.1\\
Valid MJD range  &54686-58767\\\\
Pulse frequency,~~  $f~(s^{-1}$   &29.7169027333\\
~~                                 &(0.0000148379)\\
First derivative , $ \dot{f}~(10^{-10} s^{-2} )$ &-3.71184342371 \\
~~                                  &(4.6591493867e-17)\\
Second derivative ,$ \ddot{f}~(10^{-20} s^{-3} )$ &3.3226958153\\
~~                                       &(1.1931513864e-24)\\
Third derivative , $ \dddot{f}~(10^{-28} s^{-4} )$ &1.2860657170\\
~~                                        &(9.2842295751e-32)\\
PEPOCH        &55555 \\                     
POSEPOCH       &50739\\                     
DMEPOCH        &55107.807158553\\ 
TZRMJD &56730.15526586924  \\
Solar system ephemeris model&DE405\\
Time system &TDB\\
\hline
\end{tabular}
\tablenotes{}
\end{table}
  
\subsection{Fermi-LAT}
    The LAT instrument is described by Atwood et al., (2008). We have used the already publicly available data from LAT.  A unique value of the LAT data is that a pulsar's discovery in $\gamma$-rays often enables the immediate measurement of the pulsar parameters over the ten-year span in which the LAT has been operating. LAT data have been used to find precise timing solutions for many pulsars, including radio-quiet and radio-faint pulsars (Ray et al., 2011; Kerr et al., 2015; Clark et al., 2017). 

\subsubsection{Analysis of Fermi-LAT data}
    In order to get a significant $\gamma$-ray profile to compare with the CZTI results, we took all available photon data for the LAT source PSRJ0534+2200 from MJD 57290 through 58500, bracketing the CZTI data used in this work. It leads to a total effective exposure time of 5962-kilo seconds, as the observatory scans the entire sky once every three hours. We used the HEADAS-FTOOLS \footnote{http://heasarc.gsfc.nasa.gov/ftools} on HEAsoft-ver 6.27. (Blackburn, 1995) to perform the data reduction. We obtained the Fermi-LAT data in the energy range of 0.1-300 GeV within a circular region of interest (ROI) with a 1-degree radius from the decided $\gamma$-ray position of PSR J0534+2200. We used Pass 8 data and selected events in the "Source" class (i.e. event class 2). We also excluded the events with zenith angles larger than 105 degrees to reduce the contamination by the Earth albedo $\gamma$-rays. We used $gtbary$ tool in fermi science tools to apply barycentric corrections to photon arrival times in LAT event files using corresponding Fermi orbit files. After barycentering the events using fermi science tool \footnote{https://fermi.gsfc.nasa.gov/ssc/data/analysistools/overview.html} $gtbary$, the absolute phase of each event in 0.1-300 GeV was determined by the custom code developed by us, with 6 hour SSB polycos generated using the timing parameters in Table \ref{table2}

\section{Discussion of Results}

\begin{table*}
\centering
\tabularfont
\caption{Phase component definitions for the Crab pulsar (Abdo et al., 2010)
adopted in this study}
\label{table3}
\begin{tabular}{lccc}
\topline
Component~~~~~~~~~~~~~~~~~~& Abbreviation&Phase Interbval & Width\\\midline
Peak 1&P1&0.87 - 1.07&0.20\\
Peak 2&P2&0.27 - 0.47&0.20\\
Bridge&Bridge&0.098 - 0.26&0.162\\
Off Pulse& OP&0.52 - 0.87&0.35\\
\hline
\end{tabular}
\end{table*}


\begin{figure*}
\centering
\begin{subfigure}[b]{0.53\textwidth}
                \includegraphics[width=1.6\columnwidth,height=0.85\columnwidth,angle=270]{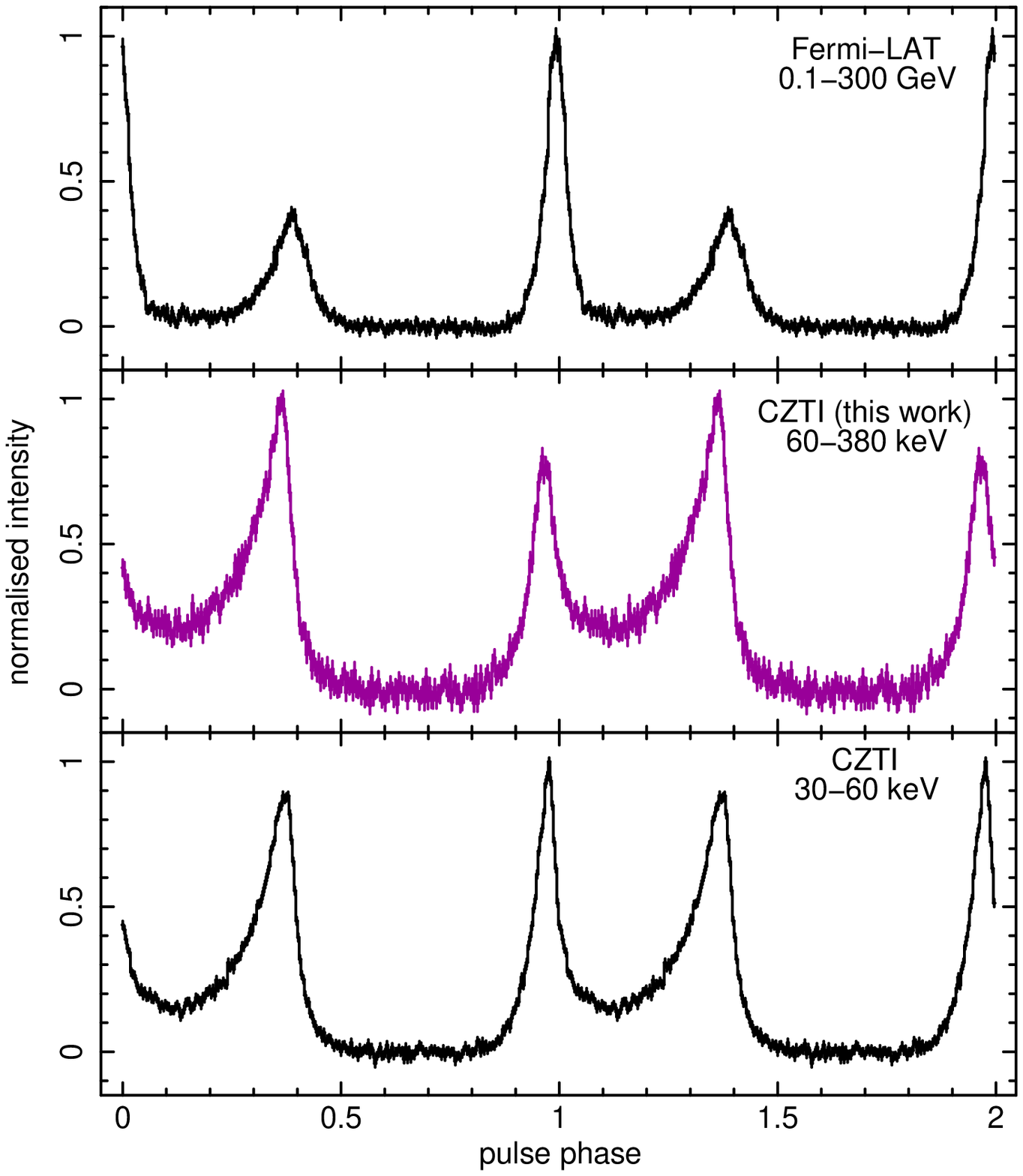}
               \caption{}
                \label{fig2(a)}
        \end{subfigure}%
        \begin{subfigure}[b]{0.53\textwidth}
                \includegraphics[width=1.6\columnwidth,height=0.85\columnwidth,angle=270]{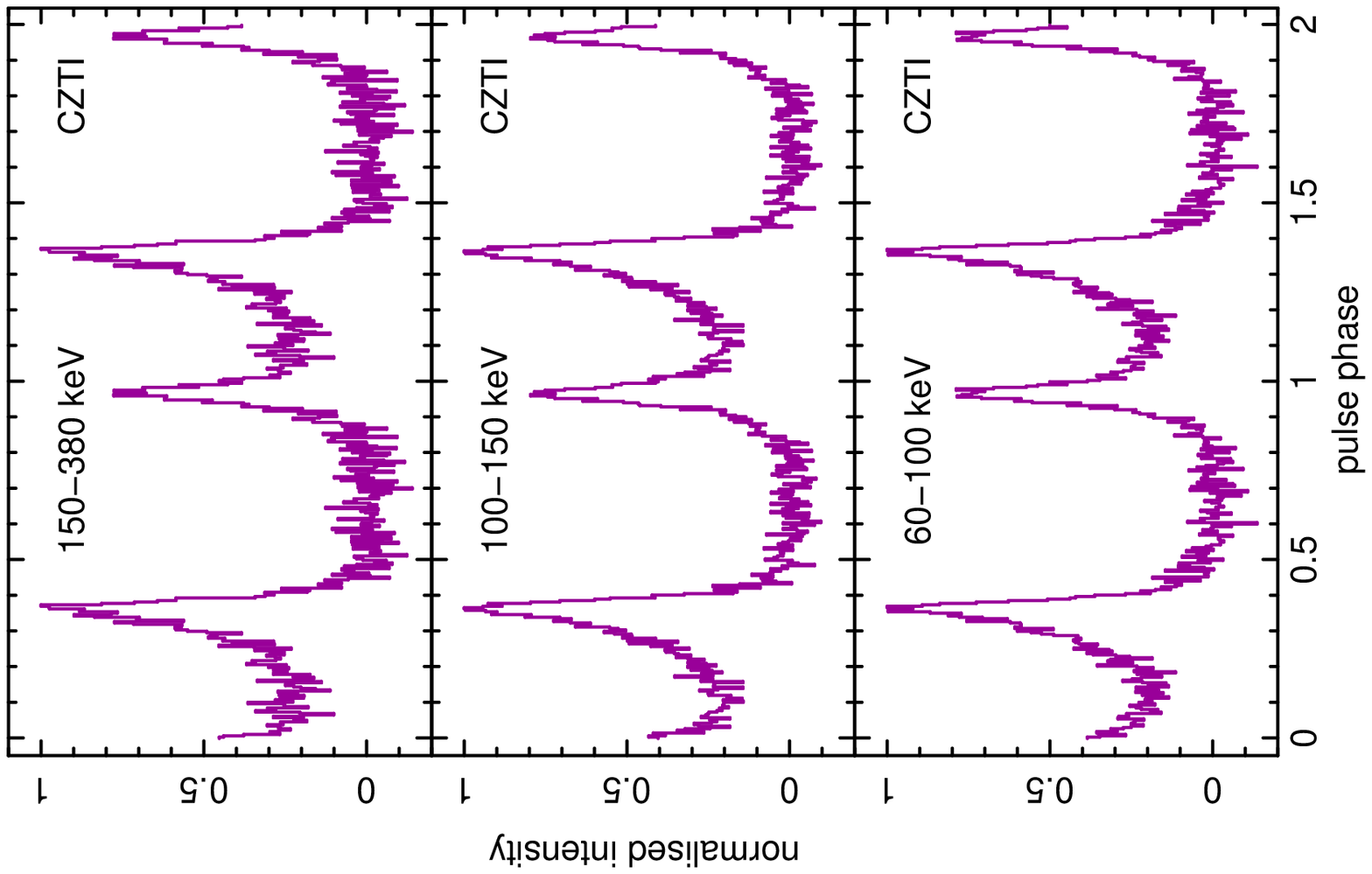}
                \caption{}
                \label{fig2(b)}
        \end{subfigure}%
        \caption{The pulse profile of the Crab pulsar observed with Fermi-LAT (black) and the AstroSat-CZTI (purple). The integration time in the 0.1--300~GeV LAT profile is $\sim$6000~ks, while that in the 60–-380~keV CZTI profiles is $\sim$5000~ks. The bottom panel of Figure \ref{fig2}(a) shows the 30-60 keV profile obtained by folding $\sim$600~ks of on-axis pointing observations of the Crab pulsar with AstroSat-CZTI. In CZTI profiles, data from multiple observations during MJD 57290-58232 with PSR J0534+2200 within 5--70 degree from pointing axis have been accumulated. Two cycles in 512 phase bins are plotted for clarity. The energy dependence of the profile is apparent here: below 60 keV the first peak P1 (near phase 1.0) dominates, and emission in the ``bridge” phase interval is moderate, while above 60 keV the second peak dominates with significant ``bridge” emission. At MeV energies, P1 starts to dominate again with strongly reduced bridge emission. A detailed reference to the energy dependence of Crab pulse profile is available in Kuiper et al., (2001).}
\label{fig2}
\end{figure*}



    Figure \ref{fig2} shows the folded pulse profile of the Crab pulsar (Period: $\sim$ 33 ms) using the LAT and the CZTI instruments. The CZTI profiles (purple in colour) accumulate data from multiple observations spread over several months, with Crab position within 5 to 70 degrees away from the nominal pointing direction. The total integration time in these CZTI profiles is about 5000~ks, giving a signal to noise ratio of $\sim 30$ in the energy integrated profile and over $\sim 15$ in the energy-resolved ones. The long-known energy dependence of the Crab pulse profile can be seen when compared with the softer bands (e.g.\ 30-60~keV in CZTI) in Figure \ref{fig2}(a). The left peak is taller at lower energies while the right peak dominates at high energies. The bridge emission connecting the two peaks is also seen relatively stronger at higher X-ray energies. The normalised light curves in three separate hard X-ray bands are shown in Figure \ref{fig2}(b). The right peak(P2) grows, but the left peak (P1) again starts to dominate at very high energies, beyond ($\sim $10 MeV) in the $\gamma$-regime, as seen in Figure \ref{fig2}(a). 
    
    For a more quantitative comment on the morphology change observed, we determined the intensity ratios P2/P1 and Bridge/P1 in three separate hard X-ray bands shown in Figure \ref{fig2}(b), adopting the phase interval definitions of (Abdo et al., 2010) shown in Table \ref{table3}. The values obtained for P2/P1 ratios are 1.301$\pm$0.002, 1.340$\pm$0.002, 1.385$\pm$0.002 and  that for Bridge/P1 are 0.536$\pm$0.002, 0.574$\pm$0.002, 0.609$\pm$0.003 in 60-100 keV, 100-150 keV and 150-380 keV bands respectively. The gradual increase of both the ratios is consistent with that reported by Kuiper et al., (2001). This validates the accuracy and stability of AstroSat-CZTI clocks and the robustness of our custom algorithm for long-term phase-connected analysis with available accurate $\gamma$-ray/radio timing models.

    Integrating the well established broken power-law model spectrum for Crab (Ulmer et al., 1995), we calculated the hard X-ray flux in 60-380 keV band to be 0.011 ph cm$^{-2}$s$^{-1}$. The observed CZTI detection count rates of  $0.876\pm 0.01$ cps in 60-380 keV band translates to an average CZTI off-axis effective area of $\sim$80 cm$^{2}$, averaged over 5-70 degree off-axis. This is about 20 percent of the on-axis effective area at lower energies.

\begin{figure}[t]
\centering
\includegraphics[width=0.8\columnwidth,height=1.4\columnwidth]{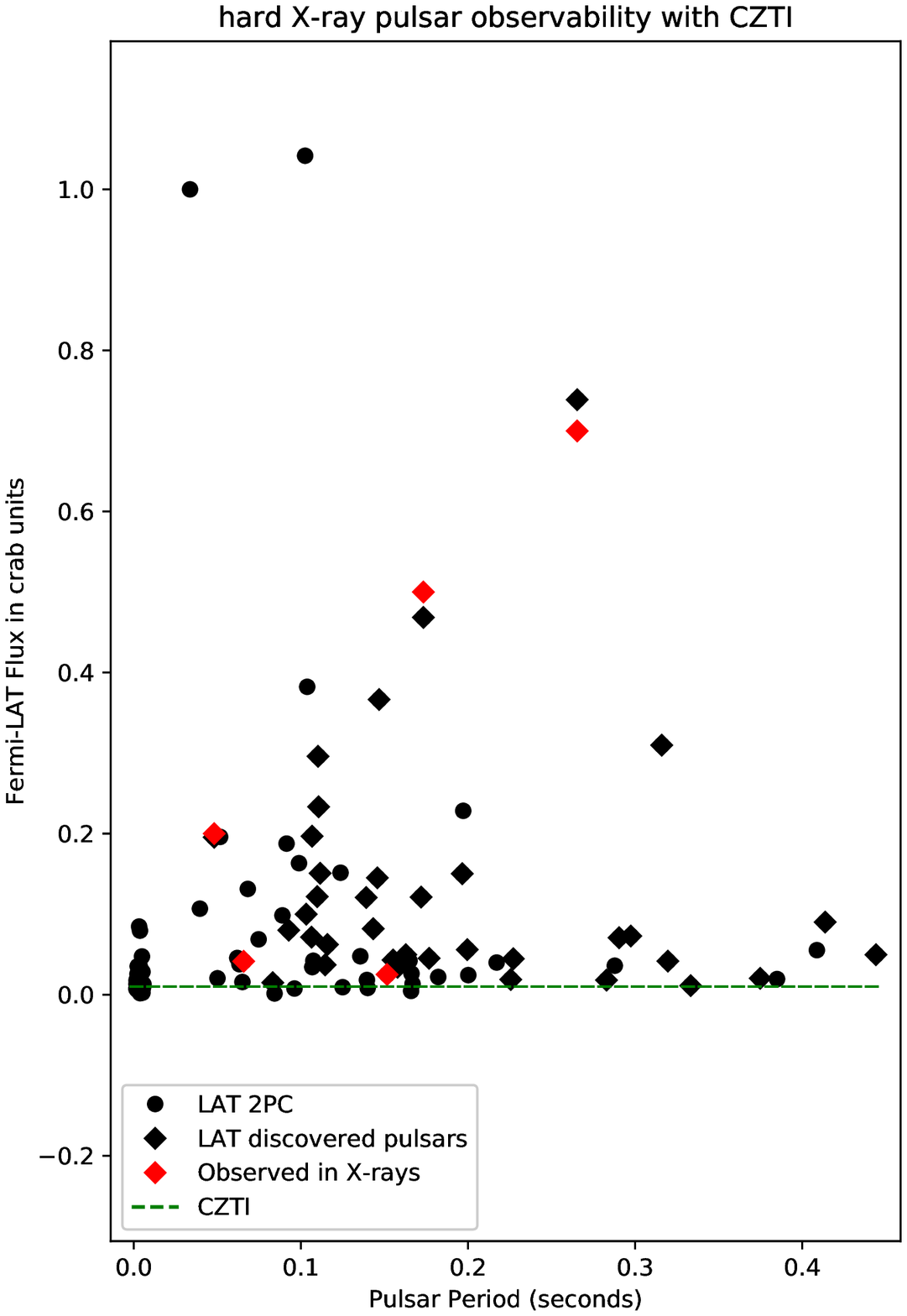}
\caption{Detectability of Fermi-LAT pulsars with the CZTI.  The green horizontal line marks the estimated flux of the faintest detectable hard X-ray pulsar using five years of archival AstroSat-CZTI data. The red diamonds represent the hard X-ray pulsar candidates discovered/observed in the soft X-ray band ($<$20 keV). The brightest $\gamma$-ray pulsars Vela (PSRJ0835-4510) and Geminga(PSRJ0633+1746), with fluxes of 7 $\&$ 3 Crab respectively, have been omitted from the figure to clearly depict the rest of the population.}
\label{fig3}
\end{figure}

    Assuming a Crab-like spectrum and based on the observed count rates mentioned above, we estimate the possible integration time required for a 5$\sigma$ detection of a 10 mCrab 
    off-axis pulsed source with CZTI to be $\sim 14$ Mega seconds. With the continuous accumulation of CZTI data since launch, one can foresee the detection of more than 90 percent of the $\gamma$-ray pulsars from the second Fermi/LAT pulsar catalog in the CZTI hard X-ray (60-380~keV) band, as shown in Figure \ref{fig3}.
 

\section{Conclusion}


We have presented a successful attempt to detect the Crab pulsar by the AstroSat CZT Imager from pointings where the pulsar was off-axis, at angles ranging from 5 to 70 degrees.  This is possible because the Collimator and the housing of the CZTI become gradually transparent at energies above 60~keV, turning it into an all-sky detector at higher energies. Accumulating an off-axis exposure of $\sim 5$~Ms, we obtain a $\sim 30\sigma$ pulse profile of the Crab pulsar in the 60--380~keV energy band.  Energy-resolved pulse profiles constructed at multiple sub-bands reproduce the known energy dependence of the profile shape.  This demonstrates the capability of AstroSat-CZTI to act as a hard X-ray pulsar monitor, in 60-380 keV band, with an average off-axis effective area of $\sim$80 cm$^{2}$. Our results establish that the CZTI time stamps possess sufficient long-term stability to carry out phase connected timing spanning many years.  We estimate that with continued accumulation of data, it will become possible for the CZTI to detect pulsars with hard X-ray fluxes down to $\sim 10$~mCrab, thus making a large majority of Fermi/LAT pulsars accessible for study in the 60--380~keV energy band.  Such a survey is currently ongoing with several successful detections already made. These results will be reported elsewhere.

\section*{Acknowledgements}

    This publication makes use of data from the CZTI onboard Indian astronomy mission AstroSat, archived at the Indian Space Science Data Centre (ISSDC). The CZT Imager instrument was built by a TIFR-led consortium of institutes across India, including VSSC, ISAC, IUCAA, SAC, and PRL. The Indian Space Research Organisation funded, managed and facilitated the project. We extend our gratitude to CZTI POC team members at IUCAA for helping with the augmentation of data. We thank Fermi Timing Observers Paul Ray and Kerr Mattew for their timely and favourable response in providing LAT ephemeris for Crab pulsar and helping with queries related to SSB polyco generation using tempo2. We thank the anonymous referee for his/her valuable suggestions to improve the paper. We would like to thank Avishek Basu, Karthik Rajeev and Atul Mohan for useful discussions. We thank IUCAA HPC facility where we carried out all the analysis. Anusree K. G. acknowledges support for this work from DST-INSPIRE Fellowship grant, IF170239, under Ministry of Science and Technology, India.


\balance



\begin{theunbibliography}{}
\vspace{-1.5em}

\bibitem{latexcompanion}
Abdo, A. A., Ackermann, M., Atwood, W. B., Bagagli, R. et al., 2009, The Astrophysical Journal, 696, 1084.

\bibitem{latexcompanion}
Abdo, A. A., et al., 2010, The Astrophysical Journal Supplement Series, 208, 17.
\bibitem{latexcompanion}
Abdo, A. A., Ackermann, M., Ajello, M. et al., 2010, The Astrophysical Journal, 708, 1254.
\bibitem{latexcompanion}
Atwood, W. B., Abdo, A. A., Ackermann, M. et al., 2009, The Astrophysical Journal, 697, 1071.
\bibitem{latexcompanion}
Bai, X. N. \& Spitkovsky, A., 2010, The Astrophysical Journal, 715, 1282
\bibitem{latexcompanion}
Basu, A., Joshi, B.C., Bhattacharya, D. et al., 2018, Astronomy and Astrophysics, 617, A22.
\bibitem{latexcompanion}
Bertsch, D. L., Brazier, K. T. S., Fichtel, C. E. et al., 1992, Nature, 357, 306.
\bibitem{latexcompanion}
 Bhalerao, V., Bhattacharya, D., Vibhute, A. et al., 2017, Journal of Astrophysics and Astronomy, 38, 31.
 \bibitem{latexcompanion}
Bhattacharya, D. (2017). Journal of Astrophysics and Astronomy, 38, 51.
\bibitem{latexcompanion}
Bhattacharya, D., Dewangan, G.C., Antia H.M.\ et al., 2018, AstroSat Handbook,  http://www.iucaa.in/\verb\~\astrosat/AstroSat\_handbook.pdf
\bibitem{latexcompanion}
Blackburn, J. K. (1995). Astronomical Data Analysis Software and Systems IV, 77, 367.

\bibitem{latextcompanion}
 Caraveo, P. A. (2014). Gamma-ray pulsar revolution. Annual Review of Astronomy and Astrophysics, 52.
 
\bibitem{latexcompanion}
Caraveo, P. A., De Luca, A., Marelli, M. et al., 2010, The Astrophysical Journal Letters, 725, L6.
\bibitem{latexcompanion}
Cerutti, B., \& Beloborodov, A. M., 2017, Space Science Reviews, 207, 111.
\bibitem{latexcompanion}
Cheng, K.S., Ho, C. \& Ruderman, M., 1986, The Astrophysical Journal, 300, 522.
\bibitem{latexcompanion}
Cheng, K.S., Ruderman, M. \& Zhang, L., 2000, The Astrophysical Journal, 537, 964.
\bibitem{latexcompanion}
Clark, C. J., Wu, J., Pletsch, H. J. et al., 2017, The Astrophysical Journal, 834, 106.
\bibitem{latexcompanion}
Du, Y., Qiao, G., \& Wang, W., 2012, The Astrophysical Journal, 748, 84.

\bibitem{latexcompanion}
Halpern, J. P., \& Holt, S. S., 1992, Nature, 357, 222.
\bibitem{latexcomapnion}
Harding, A., \& Muslimov, A., 2004, cosp, 35, 562.
\bibitem{latexcompanion}
Hewish, A., Bell, S. J., Pilkington, J. D. et al., 1968. Nature, 217, 709.

\bibitem{latexcompanion}
Hobbs, G. B., Edwards, R. T., \& Manchester, R. N., 2006, Monthly Notices of the Royal Astronomical Society, 369, 655.
\bibitem{latexcompanion}
Kerr, M., Ray, P. S., Johnston, S., 2015, The Astrophysical Journal, 814, 128.
\bibitem{latexcompanion}
Kuiper, L., Hermsen, W., Cusumano, G. et al., 2001, Astronomy \& Astrophysics, 378, 918.
\bibitem{latexcompanion}
Kuiper, L., \& Hermsen, W., 2015, Monthly Notices of the Royal Astronomical Society, 449, 3827.

\bibitem{latexcompanion}
Lin, L. C., Huang, R. H., Takata, J. et al., 2010, The Astrophysical Journal Letters, 725, L1.
\bibitem{latexcompanion}
Lin, L. C. C., Hui, C. Y., Hu, C. P. et al., 2013, The Astrophysical Journal Letters, 770, L9.

\bibitem{latexcompanion}
Lin, L. C. C., Hui, C. Y., Li, K. T. et al., 2014, The Astrophysical Journal Letters, 793, L8.
\bibitem{latexcompanion}
Marelli, M., Harding, A., Pizzocaro, D. et al., 2014, The Astrophysical Journal, 795, 168.
\bibitem{latexcompanion}
Muslimov, A.G. \& Harding, A.K., 2004, The Astrophysical Journal, 606, 1143
\bibitem{latexcompanion}
P\'etri, J., 2011, Monthly Notices of the Royal Astronomical Society, 412, 1870
\bibitem{latexcompanion}
P\'etri, J. \& Mitra, D., 2020, Monthly Notices of the Royal Astronomical Society, 491, 80.
\bibitem{latexcompanion}
Pierbattista, M., Grenier, I. A., Harding, A. K., \& Gonthier, P. L., 2012,  Astronomy \& Astrophysics, 545, A42.
\bibitem{latexcompanion}
Pierbattista, M., Grenier, I., Harding, A., \& Gonthier, P., 2014, radio and $\gamma$-ray light-curve morphology of young pulsars: comparing Fermi observations and simulated populations. Saint-Petersburg, Russia July 28–August 1, 2014, 99.
\bibitem{latexcompanion}
Ray, P. S., Kerr, M., Parent, D. et al., 2011, The Astrophysical Journal Supplement Series, 194, 17.
\bibitem{latexcompanion}
Ray, P.S., Smith D.A. et al., 2020, https://confluence.slac.stanford.edu/display/GLAMCOG/
Public+List+of+LAT-Detected+Gamma-Ray+Pulsars/

\bibitem{latexcompanion}
Romani, R. \& Yadigaroglu, I.-A., 1995, The Astrophysical Journal, 438, 314.
\bibitem{latexcompanion}
Singh, K.P., Tandon, S.N., Agrawal, P.C. et al., 2014,
SPIE, 9144E, 1S.
\bibitem{latexcompanion}
Ulmer, M. P., Matz, S. M., Grabelsky, D. A. et al., 1995, The Astrophysical Journal, 448, 356.
\bibitem{latexcompanion}
Watters, K. P., \& Romani, R. W., 2011.The Astrophysical Journal, 727, 123.
\end{theunbibliography}
\end{document}